\documentclass[pr,onecolumn,preprintnumbers,showpacs,amsmath,amssymb,showpacs,superscriptaddress,longbibliography,10pt]{revtex4-1}
\usepackage[
      colorlinks=true,
      linkcolor=blue,
      urlcolor=blue,
      filecolor=black,
      citecolor=blue,
            ]{hyperref}
\usepackage{color}
\usepackage{graphicx}
\usepackage{amsfonts}
\usepackage{amssymb}
\usepackage[margin=1in]{geometry}
\newcommand{\td}{\text{d}}

\newtheorem{corollary}{Corollary}
\newtheorem{conjecture}{Conjecture}

\begin{document}
\title{When null energy condition meets ADM mass}
\begin{abstract}
We give a conjecture on the lower bound of the ADM mass $M$ by using the null energy condition. The conjecture includes a Penrose-like inequality $3M\geq\kappa\mathcal{A}/(4\pi)+\sqrt{\mathcal{A}/4\pi}$ and the Penrose inequality $  2M\geq\sqrt{{\mathcal{A}}/{4\pi}}$ with $\mathcal{A}$ the event horizon area and  $\kappa$ the surface gravity.  Both the  conjecture in the static spherically symmetric case and the Penrose inequality for a dynamical spacetime with spherical symmetry are proved by imposing the null energy condition. We then generalize the conjecture to a general dynamical spacetime. Our results raise a new challenge for the famous unsettled question in general relativity: in what general case can the null energy condition replace other energy conditions to ensure the Penrose inequality?
\end{abstract}

\author{Run-Qiu Yang }
\email{aqiu@tju.edu.cn}
\affiliation{Center for Joint Quantum Studies and Department of Physics, School of Science, Tianjin University, Yaguan Road 135, Jinnan District, 300350 Tianjin, China}

\author{Li Li}
\email{liliphy@itp.ac.cn}
\affiliation{CAS Key Laboratory of Theoretical Physics, Institute of Theoretical Physics,
Chinese Academy of Sciences, Beijing 100190, China}
\affiliation{School of Fundamental Physics and Mathematical Sciences, Hangzhou Institute for Advanced Study, UCAS, Hangzhou 310024, China}

\author{Rong-Gen Cai}
\email{cairg@itp.ac.cn}
\affiliation{CAS Key Laboratory of Theoretical Physics, Institute of Theoretical Physics,
Chinese Academy of Sciences, Beijing 100190, China}
\affiliation{School of Fundamental Physics and Mathematical Sciences, Hangzhou Institute for Advanced Study, UCAS, Hangzhou 310024, China}

\maketitle

\noindent

\section{Motivation}
It is still an open and interesting question on the bound of mass in a given region of a spacetime. The Penrose inequality provides a lower bound for the mass of a spacetime in terms of the area of suitable surfaces that typically represent black holes. More precisely, Penrose's motivation for his inequality is as follows. Suppose one begins with asymptotically flat initial data with the ADM mass $M$, and the apparent horizon $\sigma$ with $\mathcal{A}[\sigma]$ the minimum area required to enclose $\sigma$. Evolving the system forward in time and supposing that the spacetime eventually settles down to a Kerr solution with its Bondi mass $M_f$ and the area of the event horizon $A_f$, one immediately finds $2M_f\geq\sqrt{A_f/4\pi}$ (in units of $16\pi G=c=1$). Considering the \textit{null energy condition} (NEC), the Bondi mass does not increase, while the area of the event/apparent horizon does not decrease, \emph{i.e.} $M\geq M_f$ and $A_f\geq \mathcal{A}[\sigma]$. Therefore, one obtains the ``Penrose inequality'' $2M\geq\sqrt{\mathcal{A}[\sigma]/4\pi}$. The Penrose inequality is important to gravitational collapse and the cosmic censorship conjecture. Nevertheless, finding the proof for this inequality is still a famous open problem in general relativity.
In the special case that the spacetime is static, the inequality was proved by various methods assuming weak energy condition(WEC)~\cite{Hawking1968,Geroch1973,Jang1977,Bray1997,Huisken2001,Malec:2002ki,Bray2001} (see Ref.~\cite{Mars:2009cj} for a review).  For the dynamical case, the only known proof was given for the spherically symmetric spacetime by considering the dominant energy condition (DEC) and using ``ADM energy'' instead of ADM mass~\cite{Hayward:1994bu,Mars:2009cj}. However, there are two obvious  ``gaps'': 1) only the NEC is involved in Penrose's  heuristic argument, but the current statement of the Penrose inequality requires WEC or even DEC, and 2) the spacetime is assumed to settle down to a stationary black hole in Penrose's  argument,  but this assumption plays no role in current studies and proofs. Moreover, note that the ``positive mass theorem'' is a corollary of Penrose inequality and Refs.~\cite{Penrose:1993ud,Chruciel2004} showed that the NEC can indeed insure nonnegative ADM mass. These raise  an interesting question: if assuming that the spacetime will settle down to a stationary black hole finally, is it able to prove the Penrose inequality by using the NEC only? On the other hand, the surface gravity $\kappa$ and event horizon area $ \mathcal{A}$ are two important quantities of black holes. If taking both into account, can we obtain a new lower bound on the ADM mass?

To answer these questions, we first consider the Kerr-Newman black hole with ADM mass $M$, angular momentum $Ma$, charge $Q$. We then have
\begin{equation}\label{kerrA1}
  \mathcal{A}=4\pi (r_h^2+a^2),\quad \kappa=\frac{r_h-M}{r_h^2+a^2},\quad M=\frac{r_h}{2}+\frac{a^2+Q^2}{2r_h}\,,
\end{equation}
with $r_h$ the location of the event horizon. We can obtain that
\begin{equation}\label{kerrAT1}
3M-\frac{\kappa\mathcal{A}}{4\pi}=r_h+\frac{2a^2+2Q^2}{r_h}\geq\sqrt{r_h^2+a^2}\,,
\end{equation}
and the saturation appears only if $a=Q=0$. Based on the above discussions, we propose a conjecture as follows.
\begin{conjecture}\label{theorem1}
For a 4-dimensional  asymptotically flat, static or axisymmetric stationary black hole with ``$t-\phi$'' reflection isometry, if (1) The Einstein's equation and the NEC are satisfied and (2) the cross-section of event horizon has $S^2$ topology, then there are two independent inequalities given as follows:
\begin{equation}\label{EETineq}
  3M\geq\frac{\kappa\mathcal{A}}{4\pi}+\sqrt{\frac{\mathcal{A}}{4\pi}}\,,
\end{equation}
and the Penrose inequality
\begin{equation}\label{RiePenr1}
  2M\geq\sqrt{\frac{\mathcal{A}}{4\pi}}\,.
\end{equation}
The equality is achieved only when the exterior of a black hole is Schwarzschild. Moreover, a regular stationary spacetime that is singularity free satisfies $M\geq0$ and  $M$ vanishes only for the Minkowski spacetime.
\end{conjecture}

Here we make some comments before going further. Firstly, the event horizon in \textbf{Conjecture}~\ref{theorem1} is a Killing horizon for which the surface gravity $\kappa$ is well-defined. We note that the event horizon of a stationary black hole
is not guaranteed to be a Killing horizon. According to Hawking and Ellis~\cite{hawking1}, when Einstein's equation holds with matter satisfying suitable hyperbolic equations and the DEC, the event horizon of a stationary black hole is a Killing horizon. However, here we only use the NEC. Instead, Carter proved that for a static or axisymmetric stationary black hole with ``$t-\phi$'' reflection isometry, the event horizon is a Killing horizon and the surface gravity is a constant, regardless of Einstein's equation and energy conditions~\cite{PhysRevLett.26.331}. Secondly, the surface gravity $\kappa$ depends on the normalization of the Killing vector $\xi^\mu$. This can be fixed by requiring $\xi_\mu t^\mu|_{\infty}=-1$, where $t^\mu$ is the tangent vector of the world-line of static observers at infinity. Finally, our conjecture contains a ``positive mass theorem'' as its corollary.  Though we assume static or axisymmetric stationary symmetry, we only use NEC rather than DEC or WEC. If a spacetime satisfies the NEC while breaking the WEC, the scalar curvature of the maximal slice can be negative. Then the proofs proposed by Schoen \& Yau~\cite{Schoen,PhysRevLett.43.1457} and various generalizations will lose their validity. The proofs based on spinor technique, originally proposed by Witten~\cite{Witten,Parker1982}(see also the extension to black holes~\cite{Gibbons}), require the DEC to ensure the nonnegativity of energy integration, thus would lose their validity for a spacetime that satisfies the NEC only. Compared with the results of Refs.~\cite{Penrose:1993ud,Chruciel2004}, which used the NEC to prove the nonnegativity of ADM mass, our paper offers a tighter lower bound for the ADM mass in terms of $\kappa$ and $\mathcal{A}$. Moreover, if the WEC is broken, all proofs till now about the Penrose inequality will become invalid.

\section{Proof in static spherically symmetric case}\label{Schw1}
It is clear that the Kerr-Newman black hole satisfies our \textbf{Conjecture}~\ref{theorem1}. To further support our inequalities in \textbf{Conjecture}~\ref{theorem1}, we now give proof for the static spherically symmetric case for which the metric reads
\begin{equation}\label{metric1}
  \td s^2=-f(r)e^{-\chi(r)}\td t^2+\frac{\td r^2}{f(r)}+r^2(\td\theta^2+\sin^2\theta\td\varphi^2)\,.
\end{equation}
The asymptotic flatness yields
\begin{equation}\label{asymfchi1}
  f(r)=1-\frac{2M}r+\cdots,~~|\chi(r)|\rightarrow\mathcal{O}(1/r^{1+\alpha}),~~r\rightarrow\infty\,,
\end{equation}
with $M$ the ADM mass and the constant $\alpha>0$. For a black hole, we donate the location of its event horizon to be $r_h$ at which $f(r)$ is vanishing, then the surface gravity $\kappa$ and the horizon area $\mathcal{A}$ are given by
\begin{equation}\label{ets1}
  \kappa=\frac{f'(r_h)e^{-\chi(r_h)/2}}{2},~~\mathcal{A}=4\pi r_h^2\,.
\end{equation}
For a regular case that is singularity free, both $f$ and $\chi$ are smooth at $r=0$. The energy momentum tensor ${T^\mu}_\nu$ has a form ${T^\mu}_{\nu}=\text{diag}[-\rho(r),p_r(r),p_T(r), p_T(r)]$. The Einstein's equation gives the following independent equations.
\begin{eqnarray}
  f'&=&(1-8\pi r^2\rho-f)/r\,,\label{einstf}\\
  \chi'&=&-\frac{8\pi r}f(\rho+p_r)\,,\label{einstchi}\\
p_r'&=&\frac{\rho+4p_T-3p_r}{2r}-\frac{(p_r+\rho)(8\pi p_rr^2+1)}{2rf}\,.\label{einstpr}
\end{eqnarray}
We note that the NEC insures $\chi'\leq0$. Therefore, we find from~\eqref{asymfchi1} that $\chi\geq0$.

To prove the inequalities~\eqref{EETineq} and~\eqref{RiePenr1} for the spherically symmetric case, we introduce a new ``quasi-local mass'' for an equal-$r$ surface defined as
\begin{equation}\label{defQ}
  m(r)=\frac{r^4e^{\chi/2}}6\left(\frac{fe^{-\chi}}{r^2}\right)'+\frac{r}3\,.
\end{equation}
Using Eqs.~\eqref{einstf}-\eqref{einstpr}, one can obtain that
\begin{equation}\label{diffQr}
  m'(r)=\frac{8\pi}3 e^{-\chi/2}r^2(\rho+p_T)+\frac13[1-e^{-\chi(r)/2}]\,.
\end{equation}
It is now manifest that the NEC insures $m'(r)\geq0$ outside $r_h$. Evaluating $m(r)$ at both the infinity and  the event horizon, one finds
\begin{equation}\label{Qrh1}
  M=m(\infty)\geq m(r_h)=\frac{\kappa \mathcal{A}}{12\pi}+\frac{r_h}3\,,
\end{equation}
so the inequality~\eqref{EETineq} follows. If the spacetime is regular, \emph{i.e.} horizonless and all curvature invariants are regular everywhere, we have $M=m(\infty)\geq m(0)=0$.

Next, we prove the Penrose inequality~\eqref{RiePenr1}. Solving $f(r)e^{-\chi(r)}/r^2$ in terms of $m(r)$, one obtains from~\eqref{diffQr} that
\begin{equation}\label{solufchhir}
  \frac{f(r)e^{-\chi(r)}}{r^2}=2\int_{r_h}^r\frac{3m(x)-x}{x^4}e^{-\chi(x)/2}\td x\,.
\end{equation}
Evaluating~\eqref{solufchhir} at $r\rightarrow\infty$ and noting that $3m(r)-r\leq 3M-r$ under the NEC, one has
\begin{equation}\label{solufchhir4a}
 0=\left.\frac{f(r)e^{-\chi(r)}}{2r^2}\right|_{r\rightarrow\infty}\leq\int_{r_h}^{\infty}\frac{3M-r}{r^4}e^{-\chi(r)/2}\td r\,.
\end{equation}
Monotonicity of $e^{-\chi(r)/2}$ ensures $(3M-r)[e^{-\chi(r)/2}-e^{-\chi(3M)/2}]\leq0$, which leads to $(3M-r)e^{-\chi(r)/2}\leq(3M-r)e^{-\chi(3M)/2}$. Inequality~\eqref{solufchhir4a} then implies
\begin{equation}\label{solufchhir4f}
 0\leq e^{-\chi(3M)} \int_{r_h}^{\infty}\frac{3M-r}{r^4}\td r=e^{-\chi(3M)}\left(\frac{M}{r_h^3}-\frac{1}{2r_h^2}\right)\,.
\end{equation}
One immediately finds $2M\geq r_h$, and therefore the inequality~\eqref{RiePenr1} follows.

To prove the rigidity for both inequalities~\eqref{EETineq} and \eqref{RiePenr1}, we note that they are saturated only if $\chi(r)=0$ and $m(r)=M$. This leads to $f(r)=1-2M/r$, \emph{i.e.} the Schwarzschild spacetime. Moreover, a regular stationary spacetime satisfies $M\geq0$ and $M$ vanishes only for the Minkowski spacetime.

Note that the Penrose inequality of dynamical spacetime was proved in the spherically symmetric case by considering the DEC and the ADM energy in the
literature~\cite{Hayward:1994bu,Mars:2009cj}. Interestingly, our proof for the spherically symmetric case  implies the following corollary,
\begin{corollary}\label{penrose1}
Consider a dynamical spacetime that has spherical symmetry and settles down to a static black hole finally. For an initial data set containing an apparent horizon $\sigma$, the NEC and the Einstein's equation guarantee $2M\geq\sqrt{\mathcal{A}[\sigma]/4\pi}$, where $\mathcal{A}[\sigma]$ is the minimum area required to enclose the apparent horizon $\sigma$.
\end{corollary}
The proof is as follows. Based on the Proposition 9.2.1 of Ref.~\cite{hawking1}, if the initial data set contains an apparat horizon $\sigma$ and the NEC is satisfied, there must be an event horizon $H$ and the apparent horizon lies behind $H$. We denote the intersection of $H$ and the initial data set as $\Gamma_{0}$, so $\sigma$ must be inside $\Gamma_{0}$. Since $\mathcal{A}[\sigma]$ is the minimum area required to enclose the apparent horizon, one has $\mathcal{A}[\sigma]\leq \mathcal{A}(\Gamma_{0})$. The NEC ensures that the area of the event horizon is nondecreasing, so we have  $\mathcal{A}(\Gamma_{0})\leq \mathcal{A}_\infty$, where $\mathcal{A}_\infty$ is the event horizon area at the future timelike infinity. Moreover, the mass of the final black hole is given by the Bondi mass $M_B$. Note that we have proved the Penrose inequality for a static black hole with spherical symmetry. Therefore, we have
$$2M\geq 2M_B\geq\sqrt{\mathcal{A}_{\infty}/4\pi}\geq\sqrt{\mathcal{A}(\Gamma_{0})/4\pi}\geq\sqrt{\mathcal{A}[\sigma]/4\pi}\,,$$
where we have used the fact that the Bondi mass is equal or smaller than the ADM mass.

\textbf{Corollary}~\ref{penrose1} shows that in a spherically symmetric case, the NEC is enough to ensure the Penrose inequality, the same as Penrose's heuristic argument. Compared to previous proofs in the spherically symmetric case~\cite{Mars:2009cj}, we have a natural requirement that the system will finally settle down to a static black hole. Nevertheless, our result is stronger in the following two aspects: we use the ADM mass rather than the ADM energy, and we use the NEC rather than the DEC. We also stress that $\mathcal{A}[\sigma]$ is not defined by the area of the apparent horizon $\sigma$. As pointed out by Ref.~\cite{PhysRevD.70.124031}, the apparent horizon area, in general, may not satisfy the Penrose inequality.

\section{Generalization to  dynamical black holes}
To generalize~\textbf{Conjecture}~\ref{theorem1} to the dynamical case, we should first clarify two conceptions in a non-stationary black hole: the ``horizon'' and the ``surface gravity''. Two possible candidates for the horizon in the dynamical case are the ``future outer trapping horizon'' (FOTH) introduced by Hayward~\cite{PhysRevD.49.6467} and the ``dynamical horizon'' (DH) proposed by Ashtekar~\cite{Ashtekar:2002ag}. In this paper, we will take the former.

The definition of the ``surface gravity'' in the dynamic spacetime is also a subtle issue. Once again, one has  two potential choices,   the ``trapped gravity'' proposed by Hayward~\cite{PhysRevD.49.6467} and the ``effective surface gravity'' by Ashtekar~\cite{Ashtekar:2002ag}, respectively. However,  both of them cannot reduce to the surface gravity even in the static spherically symmetric case,  Eq.~\eqref{ets1}. We now propose a new candidate of surface gravity as follows. Near the null infinities $\{\mathcal{I}_-, \mathcal{I}_+\}$ and spatial infinity $i_0$, there is an asymptotically time-like Killing vector $t^\mu$ which stands for the 4-velocity of a static observer. Take $l^\mu$ and $k^\mu$ to be, respectively, the infalling and outgoing null vectors of a FOTH. We can extend them into the whole spacetime by requiring that: 1) they are tangent vectors of null geodesics, 2) $l^\mu$ is affinely parameterized and satisfies $l_\mu t^\mu=-1$ at $\{\mathcal{I}_-, \mathcal{I}_+, i_0\}$, 3) $k^\mu$ is normalized by requiring $k^\mu l_\mu=-1$ everywhere. Then, according to the null vector fields $\{l^\mu, k^\mu$\} and their expansions $\{\theta_{(l)}, \theta_{(k)}\}$, our ``\textit{surface gravity}'' is defined as
\begin{equation}\label{defkappafoth3}
  {\kappa}=\theta_{(l)}^{-1}\mathcal{L}_l\theta_{(k)}|_{\text{FOTH}}\,.
\end{equation}
The surface gravity defined in this way is always nonnegative and will reduce to Eq.~\eqref{ets1} in the static spherically symmetric case.

We now generalize \textbf{Conjecture}~\ref{theorem1} to the dynamical case as follows.
\begin{conjecture}\label{conj2}
  For the  most outer FOTH which coincides with the event horizon at the future timelike infinity, if (1) The Einstein's equation and the NEC are satisfied, and (2) all marginal trapped surfaces of the FOTH have spherical topology, then the area $\mathcal{A}$ of a marginal trapped surface $\mathcal{S}$, the surface gravity $\kappa$, and the ADM mass of spacetime will satisfy
  \begin{equation}\label{inequeetdyna}
    3M\geq\sqrt{\mathcal{A}/(4\pi)}+\frac{1}{4\pi}\int_{\mathcal{S}}\kappa \td S\,,
  \end{equation}
  and
  \begin{equation}\label{inequeetdyna2}
    2M\geq\sqrt{\mathcal{A}/(4\pi)}\,.
  \end{equation}
If it is saturated on one marginal surface of FOTH, then the FOTH is the event horizon, and the exterior of the event horizon is Schwarzschild.
\end{conjecture}

It has been proved in Ref.~\cite{PhysRevD.49.6467} that the area of marginal surface of FOTH is nondecreasing. Therefore, one can find that the inequality~\eqref{inequeetdyna2} is a corollary of~\eqref{RiePenr1} in~\textbf{Conjecture}~\ref{theorem1}.

We now give a nontrivial check for the inequality~\eqref{inequeetdyna} by considering the generalized Vaidya solution~\cite{Wang:1998qx}:
\begin{equation}\label{vaidya2}
  \td s^2=-f(v,r)\td v^2+2\td r\td v+r^2(\td\theta^2+\sin^2\theta\td\varphi^2),\quad f(v,r)=1-2\mathcal{M}(v,r)/r\,.
\end{equation}
Note that the ADM mass $M$ is defined at the spatial infinity, \emph{i.e.} $M=\mathcal{M}(\infty,\infty)$. The marginal trapped surfaces are given by $v=const.$ and $r=const.$, thanks to the spherical symmetry. Then the infalling and outgoing null rays are, respectively, $l^\mu=(0,-1,0,0)$ and $k^\mu=(1,f/2,0,0)$ with their expansions $\theta_{(k)}=f/r$ and $\theta_{(l)}=-2/r<0$. Therefore, the FOTH is given by $f(v,r)=0$ for which we denote its solution to be $r=r_h(v)$. We then have
\begin{equation}\label{rnvaidyats}
  {\kappa}=\left.\frac{1-2\partial_r\mathcal{M}(v,r)}{2r}\right|_{r=r_h(v)},~~\mathcal{A}=4\pi r_h(v)^2,~~\mathcal{M}(v,r_h(v))=r_h(v)/2\,,
\end{equation}
from which
\begin{equation}\label{generavad0}
  3\mathcal{M}(v,r_h(v))-\frac1{4\pi}\int_{\mathcal{S}}\kappa\td S=r_h(v)+r_h(v)\partial_r\mathcal{M}(v,r)|_{r=r_h(v)}\,.
\end{equation}

The corresponding energy-momentum tensor reads~\cite{Husain:1995bf}
\begin{equation}\label{tabrn1}
  T_{\mu\nu}=ul_\mu l_\nu+2(P+\rho)l_{(\mu}k_{\nu)}+Pg_{\mu\nu}\,,
\end{equation}
with
$$u=\frac{\partial_v\mathcal{M}}{4\pi r^2},~~\rho=\frac{\partial_r\mathcal{M}}{4\pi r^2},~~P=-\frac{\partial_r^2\mathcal{M}}{8\pi r}\,.$$
This in general ($u\neq0$) describes the Type II fluids~\cite{hawking1}. The NEC demands
\begin{equation}\label{necvad1}
  u\geq0,~~\rho+P\geq0\,,
\end{equation}
as well as the following constraint on $\mathcal{M}(v,r)$.
\begin{equation}\label{necvad2a}
  \partial_v\mathcal{M}(v,r)\geq0,~~-\partial_r^2\mathcal{M}+2r^{-1}\partial_r\mathcal{M}=-r^2\partial_r(r^{-2}\partial_r\mathcal{M})\geq0\,.
\end{equation}
The asymptotic flatness requires that $\lim_{r\rightarrow\infty}r^{-2}\partial_r\mathcal{M}=0$. Therefore, the second one of Eq.~\eqref{necvad2a} implies $r^{-2}\partial_r\mathcal{M}\geq0$.  One immediately finds that $\mathcal{M}(v,r)\leq \mathcal{M}(\infty,\infty)=M$. Then, Eq.~\eqref{generavad0} shows
\begin{equation}\label{generavad}
  3M-\frac1{4\pi}\int_{\mathcal{S}}\kappa\td S\geq r_h(v)\,,
\end{equation}
so the inequality~\eqref{inequeetdyna} follows. To saturate the inequality, we need $\mathcal{M}(v,r)$ to be a constant, so the exterior is nothing but Schwarzschild. This provides nontrivial evidence to support our \textbf{Conjecture}~\ref{conj2}. Note that, in the present case, the WEC requires $\{u\geq0,\rho\geq0, P\geq0\}$ and the DEC gives $\{u\geq0,\rho\geq P\geq0\}$, both are stronger than the NEC.

\section{Summary}
To summarize, we have proposed a Penrose-like inequality involving ADM mass, surface gravity, and horizon area. For static or axisymmetric stationary black holes, our \textbf{Conjecture}~\ref{theorem1}  suggests that the Einstein's equation and the NEC ensure the Penrose-like inequality as well as the Penrose inequality. We have given a proof for the static spherically symmetric case and offered evidence for the dynamical case. In addition, the Penrose inequality for spherically symmetric (dynamic) spacetime has been proved by using the NEC rather than the DEC. Our conjecture applies in some situations not covered by previous inequalities.

Our results not only provide a new conjecture to bound the ADM mass by horizon area and surface gravity for the first time, but also raise a new challenge for the famous unsettled question in general relativity: can the NEC ensure the Penrose inequality if a spacetime settles down to a stationary black hole finally?

\begin{acknowledgments}
This work was partially supported by the National Natural Science Foundation of China Grants No.12122513, No.12075298, No.11821505, No.11991052, No.12047503, and No.12005155, and by the Key Research Program of the Chinese Academy of Sciences (CAS) Grant NO. XDPB15, the CAS Project for Young Scientists in Basic Research YSBR-006 and the Key Research Program of Frontier Sciences of CAS.
\end{acknowledgments}

%

\begin{thebibliography}{0}%
\makeatletter
\providecommand \@ifxundefined [1]{%
 \@ifx{#1\undefined}
}%
\providecommand \@ifnum [1]{%
 \ifnum #1\expandafter \@firstoftwo
 \else \expandafter \@secondoftwo
 \fi
}%
\providecommand \@ifx [1]{%
 \ifx #1\expandafter \@firstoftwo
 \else \expandafter \@secondoftwo
 \fi
}%
\providecommand \natexlab [1]{#1}%
\providecommand \enquote  [1]{``#1''}%
\providecommand \bibnamefont  [1]{#1}%
\providecommand \bibfnamefont [1]{#1}%
\providecommand \citenamefont [1]{#1}%
\providecommand \href@noop [0]{\@secondoftwo}%
\providecommand \href [0]{\begingroup \@sanitize@url \@href}%
\providecommand \@href[1]{\@@startlink{#1}\@@href}%
\providecommand \@@href[1]{\endgroup#1\@@endlink}%
\providecommand \@sanitize@url [0]{\catcode `\\12\catcode `\$12\catcode
  `\&12\catcode `\#12\catcode `\^12\catcode `\_12\catcode `\%12\relax}%
\providecommand \@@startlink[1]{}%
\providecommand \@@endlink[0]{}%
\providecommand \url  [0]{\begingroup\@sanitize@url \@url }%
\providecommand \@url [1]{\endgroup\@href {#1}{\urlprefix }}%
\providecommand \urlprefix  [0]{URL }%
\providecommand \Eprint [0]{\href }%
\providecommand \doibase [0]{http://dx.doi.org/}%
\providecommand \selectlanguage [0]{\@gobble}%
\providecommand \bibinfo  [0]{\@secondoftwo}%
\providecommand \bibfield  [0]{\@secondoftwo}%
\providecommand \translation [1]{[#1]}%
\providecommand \BibitemOpen [0]{}%
\providecommand \bibitemStop [0]{}%
\providecommand \bibitemNoStop [0]{.\EOS\space}%
\providecommand \EOS [0]{\spacefactor3000\relax}%
\providecommand \BibitemShut  [1]{\csname bibitem#1\endcsname}%
\let\auto@bib@innerbib\@empty
\end{thebibliography}%


\begin{thebibliography}{23}%
\makeatletter
\providecommand \@ifxundefined [1]{%
 \@ifx{#1\undefined}
}%
\providecommand \@ifnum [1]{%
 \ifnum #1\expandafter \@firstoftwo
 \else \expandafter \@secondoftwo
 \fi
}%
\providecommand \@ifx [1]{%
 \ifx #1\expandafter \@firstoftwo
 \else \expandafter \@secondoftwo
 \fi
}%
\providecommand \natexlab [1]{#1}%
\providecommand \enquote  [1]{``#1''}%
\providecommand \bibnamefont  [1]{#1}%
\providecommand \bibfnamefont [1]{#1}%
\providecommand \citenamefont [1]{#1}%
\providecommand \href@noop [0]{\@secondoftwo}%
\providecommand \href [0]{\begingroup \@sanitize@url \@href}%
\providecommand \@href[1]{\@@startlink{#1}\@@href}%
\providecommand \@@href[1]{\endgroup#1\@@endlink}%
\providecommand \@sanitize@url [0]{\catcode `\\12\catcode `\$12\catcode
  `\&12\catcode `\#12\catcode `\^12\catcode `\_12\catcode `\%12\relax}%
\providecommand \@@startlink[1]{}%
\providecommand \@@endlink[0]{}%
\providecommand \url  [0]{\begingroup\@sanitize@url \@url }%
\providecommand \@url [1]{\endgroup\@href {#1}{\urlprefix }}%
\providecommand \urlprefix  [0]{URL }%
\providecommand \Eprint [0]{\href }%
\providecommand \doibase [0]{http://dx.doi.org/}%
\providecommand \selectlanguage [0]{\@gobble}%
\providecommand \bibinfo  [0]{\@secondoftwo}%
\providecommand \bibfield  [0]{\@secondoftwo}%
\providecommand \translation [1]{[#1]}%
\providecommand \BibitemOpen [0]{}%
\providecommand \bibitemStop [0]{}%
\providecommand \bibitemNoStop [0]{.\EOS\space}%
\providecommand \EOS [0]{\spacefactor3000\relax}%
\providecommand \BibitemShut  [1]{\csname bibitem#1\endcsname}%
\let\auto@bib@innerbib\@empty
\bibitem [{\citenamefont {Hawking}(1968)}]{Hawking1968}%
  \BibitemOpen
  \bibfield  {author} {\bibinfo {author} {\bibfnamefont {S.~W.}\ \bibnamefont
  {Hawking}},\ }\bibfield  {title} {\enquote {\bibinfo {title} {Gravitational
  radiation in an expanding universe},}\ }\href {\doibase 10.1063/1.1664615}
  {\bibfield  {journal} {\bibinfo  {journal} {Journal of Mathematical Physics}\
  }\textbf {\bibinfo {volume} {9}},\ \bibinfo {pages} {598--604} (\bibinfo
  {year} {1968})}\BibitemShut {NoStop}%
\bibitem [{\citenamefont {Geroch}(1973)}]{Geroch1973}%
  \BibitemOpen
  \bibfield  {author} {\bibinfo {author} {\bibfnamefont {Robert}\ \bibnamefont
  {Geroch}},\ }\bibfield  {title} {\enquote {\bibinfo {title} {{ENERGY}
  {EXTRACTION}},}\ }\href {\doibase 10.1111/j.1749-6632.1973.tb41445.x}
  {\bibfield  {journal} {\bibinfo  {journal} {Annals of the New York Academy of
  Sciences}\ }\textbf {\bibinfo {volume} {224}},\ \bibinfo {pages} {108--117}
  (\bibinfo {year} {1973})}\BibitemShut {NoStop}%
\bibitem [{\citenamefont {Jang}\ and\ \citenamefont {Wald}(1977)}]{Jang1977}%
  \BibitemOpen
  \bibfield  {author} {\bibinfo {author} {\bibfnamefont {Pong~Soo}\
  \bibnamefont {Jang}}\ and\ \bibinfo {author} {\bibfnamefont {Robert~M.}\
  \bibnamefont {Wald}},\ }\bibfield  {title} {\enquote {\bibinfo {title} {The
  positive energy conjecture and the cosmic censor hypothesis},}\ }\href
  {\doibase 10.1063/1.523134} {\bibfield  {journal} {\bibinfo  {journal}
  {Journal of Mathematical Physics}\ }\textbf {\bibinfo {volume} {18}},\
  \bibinfo {pages} {41--44} (\bibinfo {year} {1977})}\BibitemShut {NoStop}%
\bibitem [{\citenamefont {Bray}(1997)}]{Bray1997}%
  \BibitemOpen
  \bibfield  {author} {\bibinfo {author} {\bibfnamefont {Hubert~L.}\
  \bibnamefont {Bray}},\ }\bibfield  {title} {\enquote {\bibinfo {title} {The
  penrose inequality in general relativity and volume comparison theorems
  involving scalar curvature},}\ }\href@noop {} {\bibfield  {journal} {\bibinfo
   {journal} {Ph.D. Thesis}\ } (\bibinfo {year} {1997})},\ \Eprint
  {http://arxiv.org/abs/0902.3241} {arXiv:0902.3241} \BibitemShut {NoStop}%
\bibitem [{\citenamefont {Huisken}\ and\ \citenamefont
  {Ilmanen}(2001)}]{Huisken2001}%
  \BibitemOpen
  \bibfield  {author} {\bibinfo {author} {\bibfnamefont {G.}~\bibnamefont
  {Huisken}}\ and\ \bibinfo {author} {\bibfnamefont {T.}~\bibnamefont
  {Ilmanen}},\ }\bibfield  {title} {\enquote {\bibinfo {title} {The inverse
  mean curvature flow and the riemannian penrose inequality},}\ }\href@noop {}
  {\bibfield  {journal} {\bibinfo  {journal} {Journal of Differential
  Geometry}\ }\textbf {\bibinfo {volume} {59}},\ \bibinfo {pages} {353--437}
  (\bibinfo {year} {2001})}\BibitemShut {NoStop}%
\bibitem [{\citenamefont {Malec}\ \emph {et~al.}(2002)\citenamefont {Malec},
  \citenamefont {Mars},\ and\ \citenamefont {Simon}}]{Malec:2002ki}%
  \BibitemOpen
  \bibfield  {author} {\bibinfo {author} {\bibfnamefont {Edward}\ \bibnamefont
  {Malec}}, \bibinfo {author} {\bibfnamefont {Marc}\ \bibnamefont {Mars}}, \
  and\ \bibinfo {author} {\bibfnamefont {Walter}\ \bibnamefont {Simon}},\
  }\bibfield  {title} {\enquote {\bibinfo {title} {{On the Penrose inequality
  for general horizons}},}\ }\href {\doibase 10.1103/PhysRevLett.88.121102}
  {\bibfield  {journal} {\bibinfo  {journal} {Phys. Rev. Lett.}\ }\textbf
  {\bibinfo {volume} {88}},\ \bibinfo {pages} {121102} (\bibinfo {year}
  {2002})},\ \Eprint {http://arxiv.org/abs/gr-qc/0201024} {arXiv:gr-qc/0201024}
  \BibitemShut {NoStop}%
\bibitem [{\citenamefont {Bray}(2001)}]{Bray2001}%
  \BibitemOpen
  \bibfield  {author} {\bibinfo {author} {\bibfnamefont {Hubert~L.}\
  \bibnamefont {Bray}},\ }\bibfield  {title} {\enquote {\bibinfo {title} {Proof
  of the riemannian penrose inequality using the positive mass theorem},}\
  }\href {\doibase 10.4310/jdg/1090349428} {\bibfield  {journal} {\bibinfo
  {journal} {Journal of Differential Geometry}\ }\textbf {\bibinfo {volume}
  {59}} (\bibinfo {year} {2001}),\ 10.4310/jdg/1090349428}\BibitemShut
  {NoStop}%
\bibitem [{\citenamefont {Mars}(2009)}]{Mars:2009cj}%
  \BibitemOpen
  \bibfield  {author} {\bibinfo {author} {\bibfnamefont {Marc}\ \bibnamefont
  {Mars}},\ }\bibfield  {title} {\enquote {\bibinfo {title} {{Present status of
  the Penrose inequality}},}\ }\href {\doibase 10.1088/0264-9381/26/19/193001}
  {\bibfield  {journal} {\bibinfo  {journal} {Class. Quant. Grav.}\ }\textbf
  {\bibinfo {volume} {26}},\ \bibinfo {pages} {193001} (\bibinfo {year}
  {2009})},\ \Eprint {http://arxiv.org/abs/0906.5566} {arXiv:0906.5566 [gr-qc]}
  \BibitemShut {NoStop}%
\bibitem [{\citenamefont {Hayward}(1996)}]{Hayward:1994bu}%
  \BibitemOpen
  \bibfield  {author} {\bibinfo {author} {\bibfnamefont {Sean~A.}\ \bibnamefont
  {Hayward}},\ }\bibfield  {title} {\enquote {\bibinfo {title} {{Gravitational
  energy in spherical symmetry}},}\ }\href {\doibase 10.1103/PhysRevD.53.1938}
  {\bibfield  {journal} {\bibinfo  {journal} {Phys. Rev. D}\ }\textbf {\bibinfo
  {volume} {53}},\ \bibinfo {pages} {1938--1949} (\bibinfo {year} {1996})},\
  \Eprint {http://arxiv.org/abs/gr-qc/9408002} {arXiv:gr-qc/9408002}
  \BibitemShut {NoStop}%
\bibitem [{\citenamefont {Penrose}\ \emph {et~al.}(1993)\citenamefont
  {Penrose}, \citenamefont {Sorkin},\ and\ \citenamefont
  {Woolgar}}]{Penrose:1993ud}%
  \BibitemOpen
  \bibfield  {author} {\bibinfo {author} {\bibfnamefont {R.}~\bibnamefont
  {Penrose}}, \bibinfo {author} {\bibfnamefont {R.~D.}\ \bibnamefont {Sorkin}},
  \ and\ \bibinfo {author} {\bibfnamefont {E.}~\bibnamefont {Woolgar}},\
  }\bibfield  {title} {\enquote {\bibinfo {title} {{A Positive mass theorem
  based on the focusing and retardation of null geodesics}},}\ }\href@noop {}
  {\  (\bibinfo {year} {1993})},\ \Eprint {http://arxiv.org/abs/gr-qc/9301015}
  {arXiv:gr-qc/9301015} \BibitemShut {NoStop}%
\bibitem [{\citenamefont {Chru\'{s}ciel}(2004)}]{Chruciel2004}%
  \BibitemOpen
  \bibfield  {author} {\bibinfo {author} {\bibfnamefont {Piotr~T}\ \bibnamefont
  {Chru\'{s}ciel}},\ }\bibfield  {title} {\enquote {\bibinfo {title} {A poor
  man's positive energy theorem: {II}. null geodesics},}\ }\href {\doibase
  10.1088/0264-9381/21/18/008} {\bibfield  {journal} {\bibinfo  {journal}
  {Classical and Quantum Gravity}\ }\textbf {\bibinfo {volume} {21}},\ \bibinfo
  {pages} {4399--4415} (\bibinfo {year} {2004})}\BibitemShut {NoStop}%
\bibitem [{\citenamefont {Hawking}\ and\ \citenamefont
  {Ellis}(1973)}]{hawking1}%
  \BibitemOpen
  \bibfield  {author} {\bibinfo {author} {\bibfnamefont {S.~W.}\ \bibnamefont
  {Hawking}}\ and\ \bibinfo {author} {\bibfnamefont {G.~F.~R.}\ \bibnamefont
  {Ellis}},\ }\href@noop {} {\emph {\bibinfo {title} {The Large Scale Structure
  of Space-Time}}}\ (\bibinfo  {publisher} {Cambridge University Press},\
  \bibinfo {address} {New York},\ \bibinfo {year} {1973})\BibitemShut {NoStop}%
\bibitem [{\citenamefont {Carter}(1971)}]{PhysRevLett.26.331}%
  \BibitemOpen
  \bibfield  {author} {\bibinfo {author} {\bibfnamefont {B.}~\bibnamefont
  {Carter}},\ }\bibfield  {title} {\enquote {\bibinfo {title} {Axisymmetric
  black hole has only two degrees of freedom},}\ }\href {\doibase
  10.1103/PhysRevLett.26.331} {\bibfield  {journal} {\bibinfo  {journal} {Phys.
  Rev. Lett.}\ }\textbf {\bibinfo {volume} {26}},\ \bibinfo {pages} {331--333}
  (\bibinfo {year} {1971})}\BibitemShut {NoStop}%
\bibitem [{\citenamefont {Schoen}\ and\ \citenamefont
  {Yau}(1979{\natexlab{a}})}]{Schoen}%
  \BibitemOpen
  \bibfield  {author} {\bibinfo {author} {\bibfnamefont {Richard}\ \bibnamefont
  {Schoen}}\ and\ \bibinfo {author} {\bibfnamefont {Shing-Tung}\ \bibnamefont
  {Yau}},\ }\bibfield  {title} {\enquote {\bibinfo {title} {{On the proof of
  the positive mass conjecture in general relativity}},}\ }\href {\doibase
  10.1007/BF01940959} {\bibfield  {journal} {\bibinfo  {journal} {Commun.Math.
  Phys}\ }\textbf {\bibinfo {volume} {65}},\ \bibinfo {pages} {45--76}
  (\bibinfo {year} {1979}{\natexlab{a}})}\BibitemShut {NoStop}%
\bibitem [{\citenamefont {Schoen}\ and\ \citenamefont
  {Yau}(1979{\natexlab{b}})}]{PhysRevLett.43.1457}%
  \BibitemOpen
  \bibfield  {author} {\bibinfo {author} {\bibfnamefont {Richard}\ \bibnamefont
  {Schoen}}\ and\ \bibinfo {author} {\bibfnamefont {Shing-Tung}\ \bibnamefont
  {Yau}},\ }\bibfield  {title} {\enquote {\bibinfo {title} {Positivity of the
  total mass of a general space-time},}\ }\href {\doibase
  10.1103/PhysRevLett.43.1457} {\bibfield  {journal} {\bibinfo  {journal}
  {Phys. Rev. Lett.}\ }\textbf {\bibinfo {volume} {43}},\ \bibinfo {pages}
  {1457--1459} (\bibinfo {year} {1979}{\natexlab{b}})}\BibitemShut {NoStop}%
\bibitem [{\citenamefont {Witten}(1981)}]{Witten}%
  \BibitemOpen
  \bibfield  {author} {\bibinfo {author} {\bibfnamefont {E.}~\bibnamefont
  {Witten}},\ }\bibfield  {title} {\enquote {\bibinfo {title} {{A new proof of
  the positive energy theorem}},}\ }\href {\doibase 10.1007/BF01208277}
  {\bibfield  {journal} {\bibinfo  {journal} {Commun.Math. Phys}\ }\textbf
  {\bibinfo {volume} {80}},\ \bibinfo {pages} {381--402} (\bibinfo {year}
  {1981})}\BibitemShut {NoStop}%
\bibitem [{\citenamefont {Parker}\ and\ \citenamefont
  {Taubes}(1982)}]{Parker1982}%
  \BibitemOpen
  \bibfield  {author} {\bibinfo {author} {\bibfnamefont {Thomas}\ \bibnamefont
  {Parker}}\ and\ \bibinfo {author} {\bibfnamefont {Clifford~Henry}\
  \bibnamefont {Taubes}},\ }\bibfield  {title} {\enquote {\bibinfo {title} {{On
  Witten's proof of the positive energy theorem}},}\ }\href {\doibase
  10.1007/bf01208569} {\bibfield  {journal} {\bibinfo  {journal}
  {Communications in Mathematical Physics}\ }\textbf {\bibinfo {volume} {84}},\
  \bibinfo {pages} {223--238} (\bibinfo {year} {1982})}\BibitemShut {NoStop}%
\bibitem [{\citenamefont {Gibbons}\ \emph {et~al.}(1983)\citenamefont
  {Gibbons}, \citenamefont {Hawking}, \citenamefont {Horowitz},\ and\
  \citenamefont {Perry}}]{Gibbons}%
  \BibitemOpen
  \bibfield  {author} {\bibinfo {author} {\bibfnamefont {G.~W.}\ \bibnamefont
  {Gibbons}}, \bibinfo {author} {\bibfnamefont {S.~W.}\ \bibnamefont
  {Hawking}}, \bibinfo {author} {\bibfnamefont {Gary~T.}\ \bibnamefont
  {Horowitz}}, \ and\ \bibinfo {author} {\bibfnamefont {Malcolm~J.}\
  \bibnamefont {Perry}},\ }\bibfield  {title} {\enquote {\bibinfo {title}
  {{Positive mass theorems for black holes}},}\ }\href {\doibase
  10.1007/BF01213209} {\bibfield  {journal} {\bibinfo  {journal} {Commun.Math.
  Phys}\ }\textbf {\bibinfo {volume} {88}},\ \bibinfo {pages} {295--308}
  (\bibinfo {year} {1983})}\BibitemShut {NoStop}%
\bibitem [{\citenamefont {Ben-Dov}(2004)}]{PhysRevD.70.124031}%
  \BibitemOpen
  \bibfield  {author} {\bibinfo {author} {\bibfnamefont {Ishai}\ \bibnamefont
  {Ben-Dov}},\ }\bibfield  {title} {\enquote {\bibinfo {title} {Penrose
  inequality and apparent horizons},}\ }\href {\doibase
  10.1103/PhysRevD.70.124031} {\bibfield  {journal} {\bibinfo  {journal} {Phys.
  Rev. D}\ }\textbf {\bibinfo {volume} {70}},\ \bibinfo {pages} {124031}
  (\bibinfo {year} {2004})}\BibitemShut {NoStop}%
\bibitem [{\citenamefont {Hayward}(1994)}]{PhysRevD.49.6467}%
  \BibitemOpen
  \bibfield  {author} {\bibinfo {author} {\bibfnamefont {Sean~A.}\ \bibnamefont
  {Hayward}},\ }\bibfield  {title} {\enquote {\bibinfo {title} {General laws of
  black-hole dynamics},}\ }\href {\doibase 10.1103/PhysRevD.49.6467} {\bibfield
   {journal} {\bibinfo  {journal} {Phys. Rev. D}\ }\textbf {\bibinfo {volume}
  {49}},\ \bibinfo {pages} {6467--6474} (\bibinfo {year} {1994})},\ \Eprint
  {http://arxiv.org/abs/gr-qc/9303006} {arXiv:gr-qc/9303006} \BibitemShut
  {NoStop}%
\bibitem [{\citenamefont {Ashtekar}\ and\ \citenamefont
  {Krishnan}(2002)}]{Ashtekar:2002ag}%
  \BibitemOpen
  \bibfield  {author} {\bibinfo {author} {\bibfnamefont {Abhay}\ \bibnamefont
  {Ashtekar}}\ and\ \bibinfo {author} {\bibfnamefont {Badri}\ \bibnamefont
  {Krishnan}},\ }\bibfield  {title} {\enquote {\bibinfo {title} {{Dynamical
  horizons: Energy, angular momentum, fluxes and balance laws}},}\ }\href
  {\doibase 10.1103/PhysRevLett.89.261101} {\bibfield  {journal} {\bibinfo
  {journal} {Phys. Rev. Lett.}\ }\textbf {\bibinfo {volume} {89}},\ \bibinfo
  {pages} {261101} (\bibinfo {year} {2002})},\ \Eprint
  {http://arxiv.org/abs/gr-qc/0207080} {arXiv:gr-qc/0207080} \BibitemShut
  {NoStop}%
\bibitem [{\citenamefont {Wang}\ and\ \citenamefont {Wu}(1999)}]{Wang:1998qx}%
  \BibitemOpen
  \bibfield  {author} {\bibinfo {author} {\bibfnamefont {Anzhong}\ \bibnamefont
  {Wang}}\ and\ \bibinfo {author} {\bibfnamefont {Yumei}\ \bibnamefont {Wu}},\
  }\bibfield  {title} {\enquote {\bibinfo {title} {{Generalized Vaidya
  solutions}},}\ }\href {\doibase 10.1023/A:1018819521971} {\bibfield
  {journal} {\bibinfo  {journal} {Gen. Rel. Grav.}\ }\textbf {\bibinfo {volume}
  {31}},\ \bibinfo {pages} {107} (\bibinfo {year} {1999})},\ \Eprint
  {http://arxiv.org/abs/gr-qc/9803038} {arXiv:gr-qc/9803038} \BibitemShut
  {NoStop}%
\bibitem [{\citenamefont {Husain}(1996)}]{Husain:1995bf}%
  \BibitemOpen
  \bibfield  {author} {\bibinfo {author} {\bibfnamefont {Viqar}\ \bibnamefont
  {Husain}},\ }\bibfield  {title} {\enquote {\bibinfo {title} {{Exact solutions
  for null fluid collapse}},}\ }\href {\doibase 10.1103/PhysRevD.53.R1759}
  {\bibfield  {journal} {\bibinfo  {journal} {Phys. Rev. D}\ }\textbf {\bibinfo
  {volume} {53}},\ \bibinfo {pages} {1759--1762} (\bibinfo {year} {1996})},\
  \Eprint {http://arxiv.org/abs/gr-qc/9511011} {arXiv:gr-qc/9511011}
  \BibitemShut {NoStop}%
\end{thebibliography}
%

\end{document}